\documentclass[a4paper,11pt]{article}
\usepackage{jcappub} 
\usepackage{indentfirst}
\usepackage{epsf,amsmath,graphicx,amssymb,url,hyperref,dcolumn}
\usepackage{subcaption}
\usepackage{mathtools}
\usepackage{color}
\usepackage{longtable,tabu}
\usepackage[normalem]{ulem}
\usepackage{adjustbox}
\usepackage{array}
\usepackage{xcolor}

\title{Directional Tests of the Cosmic Distance Duality Relation using Pantheon+ and BAO}
\author[a]{Pushty Shrimankar}
\author[a]{and Gopal Kashyap\footnote{Corresponding author: gopal.hari@vit.ac.in}}
\emailAdd{pushty.shrimankar2024@vitstudent.ac.in}
\emailAdd{gopal.hari@vit.ac.in}

\affiliation[a]{ Department of Physics, School of Advanced Sciences,
Vellore Institute of Technology, Vellore, Tamil Nadu 632014, India }

\abstract{
We present a model-independent test of anisotropy in the cosmic distance duality relation (CDDR), $D_L=(1+z)^2 D_A$, using the Pantheon+ type Ia supernova sample and baryon acoustic oscillation (BAO) data. The angular diameter distance is reconstructed via Gaussian Processes, enabling an estimate of $\eta(z)=D_L/[D_A(1+z)^2]$ without assuming a background cosmology. We also allow for a possible isotropic evolution, parameterized as $\eta(z)=1+\eta_1 z$, and find a redshift-dependent deviation whose significance depends on the assumed supernova calibration.
Anisotropy is modeled through a dipole modulation and constrained using a full covariance-based likelihood. To assess statistical significance, we construct null realizations that preserve both the redshift distribution and the survey selection function. We find that the observed dipole amplitude is consistent with isotropic expectations and lies below the levels induced by statistical fluctuations and survey geometry. We obtain a robust 95\% upper bound $A_{95}=0.025$, stable across different supernova calibration choices.
We find no evidence for intrinsic anisotropy in the CDDR. Our results highlight the importance of accounting for survey selection effects in anisotropy searches and provide a viable framework for testing directional deviations in cosmological relations.
}

\begin{document}
\maketitle
\section{Introduction}

The Cosmic Distance Duality Relation (CDDR), also known as the Etherington reciprocity relation \cite{Etherington1933,Etherington2007}, establishes a fundamental connection between two key cosmological distance measures, the luminosity distance, $D_L$, and the angular diameter distance, $D_A$. At a given redshift $z$, it is defined as
\begin{equation}
\label{eq:CDDR_def}
\eta(z) \equiv \frac{D_L(z)}{(1+z)^2 D_A(z)} = 1,
\end{equation}
or equivalently $D_L(z) = (1+z)^2 D_A(z)$. This relation is purely geometrical in origin and independent of the dynamical evolution of the universe. As first demonstrated by Etherington and later formalized within relativistic cosmology \cite{1971grc..conf..104E}, it follows directly from photon propagation along null geodesics and the conservation of photon number in a general curved spacetime.

Any departure from the CDDR would hint at a failure of one or more of these essential assumptions, potentially revealing new physics. Some possible reasons for these violations are photon interactions with exotic particles \cite{Bassett2004}, absorption or scattering by cosmic dust \cite{Corasaniti_2006}, changes in fundamental constants \cite{Ellis2013}, and axion-photon conversion \cite{Tiwari_2017,manuel2022}. So, the CDDR has undergone extensive validation through various observational probes, mostly by using luminosity distances from type Ia supernovae (SNe Ia) with angular diameter distances obtained from galaxy clusters, strong gravitational lensing, or baryon acoustic oscillations \cite{Uzan2004,Holanda_2010,Holanda_2016,Liao_2016,xu_model-independent_2022, zhang2025, Li2025}.

All cosmological observations are interpreted within the standard cosmological model,$\Lambda$CDM, which rests on the assumption of statistical homogeneity and isotropy on large scales, known as the "Cosmological Principle". This assumption is strongly supported by observations of the cosmic microwave background (CMB), especially the near-isotropy of temperature fluctuations measured by the Wilkinson Microwave Anisotropy Probe and the Planck satellite \cite{Bennett2013,planck2018}. In this framework, the dominant CMB dipole is understood as a kinematic effect, produced by the motion of the observer relative to the cosmic rest frame.

If this idea is correct, the same motion should also appear in how galaxies and quasars are spread across the sky. But radio surveys like NVSS see a dipole that is much larger than expected from the CMB alone \cite{Blake_2002,Singal_2011,Tiwari_2015,Domenech_2022,kashyap2025}. This discrepancy is known as the "cosmic dipole tension". It has attracted much attention because it hints at something beyond a purely kinematic origin. More recently, supernova data have also suggested possible dipolar patterns in the expansion rate of the universe \cite{Sah2025,Yoo_2025}. Taken together, these observations provide strong motivation to search carefully for directional effects in cosmological data.

There are also other hints of large‑scale anisotropy. For example, some studies find alignments in light polarisation from stars and galaxies \cite{Hutsemekers_1998,Hutsemekers_Lamy_2001,Tiwari_Jain_2013,Pelgrims_Hutsemekers_2015}, or uneven distributions of radio sources and jet directions \cite{Taylor,Panwar}. Others report large‑scale flows of matter \cite{Kashlinsky_2008}, or a north-south asymmetry in the CMB \cite{Eriksen_2004_hemispherical_power_asymmetry,Quadrupole_and_octopole_alignment,Octo_Quadrupole_align}. 

If such a large‑scale anisotropy really exists, it should also affect the distance duality relation. In other words, distances measured toward different parts of the sky would no longer agree with each other. From a theoretical standpoint, the validity of the CDDR is closely tied to the geometry of spacetime. It holds in Riemannian geometry \cite{Etherington2007}, and therefore it is preserved in standard metric theories of gravity, even in anisotropic ones like Bianchi models. However, violations can appear if photons interact with exotic particles or if we go beyond Riemannian geometry. This makes the CDDR a particularly sensitive probe of both new physics and possible departures from the cosmological principle. Moreover, even if the CDDR holds at each point in the sky, averaging over directions can still hide directional variations in the expansion history. Hence, a  dedicated direction dependent test for CDDR is essential.

Despite this, most existing tests of the CDDR assume statistical isotropy. They compare distance measures derived from different observables without accounting for possible directional dependence. If the universe truly exhibits anisotropic features, such comparisons may be biased. The reason is that distances inferred from objects located in different regions of the sky are not necessarily directly comparable. This motivates a generalized formulation in which the CDDR depends explicitly on both redshift and direction, which we follow in this work.

Anisotropic extensions of the CDDR have also been studied in Ref.~\cite{Li_2018}. In that work, the authors applied a dipole parametrization by comparing luminosity distances from type Ia supernovae (using the Union 2.1 and JLA compilations) with angular diameter distances from strong gravitational lensing systems at the same redshifts. They found no significant deviation from the standard relation. They also performed Monte Carlo simulations with synthetic datasets to validate their dipole estimators. To test their method, they created simulated datasets, some with an artificial dipole and some without, to check whether their approach could correctly recover an injected signal and to understand its behaviour under idealised conditions.

In the present work, we adopt a more robust and comprehensive approach. We use the recent Pantheon+ supernova compilation, which includes full statistical and systematic covariances, and combine it with a modern BAO dataset that incorporates 6dFGS, SDSS, BOSS, eBOSS, and DESI measurements. Instead of relying on strong lensing, we reconstruct the BAO‑derived luminosity distance using Gaussian processes. This model‑independent interpolation allows us to estimate $\eta(z)$ at any supernova redshift without assuming a specific cosmological background. Furthermore, we treat the absolute magnitude $M_B$ with different priors (uniform, DB23, SH0ES) to demonstrate that the anisotropy constraints are calibration‑independent.

To assess the statistical significance of anisotropy in real observational data, we construct null realisations that preserve both the redshift distribution and the actual survey selection function. This approach incorporates the effects of non‑uniform sky coverage and the realistic covariance structure. It enables a direct comparison between the observed dipole amplitude and the distribution expected under isotropy in the presence of survey systematics. Our null tests explicitly account for the survey geometry, providing a more realistic and robust evaluation of any potential dipole signal.

In our analysis, the supernova data provide most of the angular information. The BAO data, however, anchor the angular diameter distance, so that any anisotropy we look for is in the relation between $D_L$ and $D_A$, not just in $D_L$ alone. In other words, we are testing the CDDR itself for directional dependence. By combining supernovae and BAO in a largely model‑independent way, we reconstruct $\eta(z,\hat{n})$ and at the same time constrain how it changes with redshift and with direction. This gives us a rigorous, all‑around test of the basic assumptions underlying standard cosmology.

This paper is organised as follows:

\section{Methodology}

\subsection{Anisotropic parameterization}
\label{sec:anisotropic_estimator}

To probe possible violations of statistical isotropy in the distance duality relation (DDR), we extend the standard parametrization to include directional dependence. In general, the DDR can be written as a function of redshift $z$ and sky direction $\hat{n}$ as
\begin{equation}
\eta(z,\hat{n}) = \frac{D_L(z,\hat{n})}{(1+z)^2 D_A(z,\hat{n})} , .
\end{equation}
For small anisotropic deviations, the angular dependence can be expanded in spherical harmonics,
\begin{equation}
\eta(z,\hat{n}) = \eta_0(z) \left[1 + \sum_{\ell m} a_{\ell m}(z), Y_{\ell m}(\hat{n}) \right],
\end{equation}
where $\eta_0(z)$ represents the isotropic component and the coefficients $a_{\ell m}(z)$ quantify directional deviations.
We restrict our analysis to the dipole ($\ell=1$) component, which is expected to provide the leading contribution to large-scale anisotropy. Higher-order multipoles are increasingly difficult to constrain with the current sky coverage and data volume. So the expansion reduces to
\begin{equation}
\eta(z,\hat{n}) = \eta_0(z)\left[1 + \hat{n}\cdot \mathbf{D}(z)\right],
\end{equation}
where $\mathbf{D}(z)$ is the dipole vector. In this work, we adopt a simple linear parameterization for the isotropic evolution (see e.g~\cite{xu_model-independent_2022,Li2025,zhang2025}),
\begin{equation}
\eta_0(z) = 1 + \eta_1 z,
\end{equation}
such that the full model becomes
\begin{equation}
\eta(z,\hat{n}) = (1 + \eta_1 z)\left(1 + \hat{n}\cdot \mathbf{D}\right).
\end{equation}

Given a set of $N$ observations $\eta_i \equiv \eta(z_i,\hat{n}_i)$ with covariance matrix $\mathbf{C}$, the model parameters can be inferred by minimizing the quadratic form
\begin{equation}
\chi^2 = (\boldsymbol{\eta} - \boldsymbol{\eta}_{\rm model})^{\rm T}\,\mathbf{C}^{-1}\,(\boldsymbol{\eta} - \boldsymbol{\eta}_{\rm model}) .
\end{equation}
In this work we determine $\mathbf{D}$ jointly with $\eta_1$ and other nuisance parameters through a full likelihood analysis, allowing for a consistent propagation of uncertainties.

\subsection{Data Sets}

Our analysis combines three types of cosmological observations, luminosity distances from Type Ia supernovae, angular diameter distances inferred from baryon acoustic oscillations (BAO), and Hubble parameter measurements from cosmic chronometers (CC). The latter are used to reconstruct the expansion history $H(z)$ in a model-independent way, which is necessary to convert the isotropic BAO quantity $D_V/r_d$ into the angular diameter distance $D_A(z)$.

\begin{enumerate}

\item \textbf{Pantheon+ Supernovae Sample:}
We use the Pantheon+ compilation of 1701 Type Ia supernovae\footnote{\url{https://github.com/CobayaSampler/sn_data}} \cite{Scolnic_2022}. The observed luminosity distance is derived directly from the distance modulus,
\begin{equation}
D_L^{\mathrm{SN}}(z) = 10^{\frac{\mu(z) - 25}{5}} , \mathrm{Mpc},
\end{equation}
where $\mu = m_B - M_B$ and $M_B$ is treated as a free nuisance parameter.

To avoid extrapolation of the BAO Gaussian process reconstruction and to reduce the influence of peculiar velocities, we restrict the supernova sample to $z > 0.1$. This cut leaves 960 SNe, providing sufficient statistical power to constrain a dipole of the scale suggested by recent large-scale anisotropy hints. The angular distribution of these SNe in Galactic coordinates is shown in Fig.~\ref{fig:dist}(a), while their redshift distribution is presented in Fig.~\ref{fig:dist}(b). 

\begin{figure}[ht]
    \centering
    \begin{subfigure}{0.48\textwidth}
        \centering
        \includegraphics[width=\linewidth]{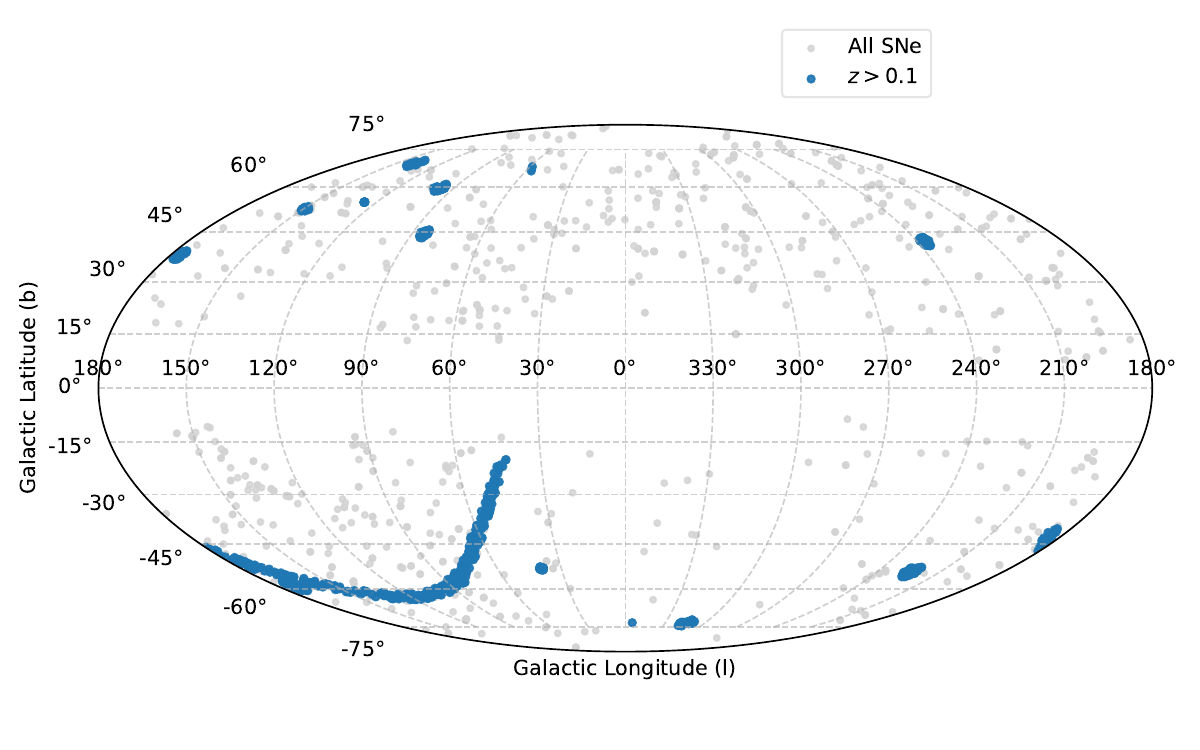}
        \caption{Mollweide projection of the 960 SNe with $z>0.1$ used in this work.}
        \label{fig:mollweide_galactic}
    \end{subfigure}
    \hfill
    \begin{subfigure}{0.48\textwidth}
        \centering
        \includegraphics[width=\linewidth]{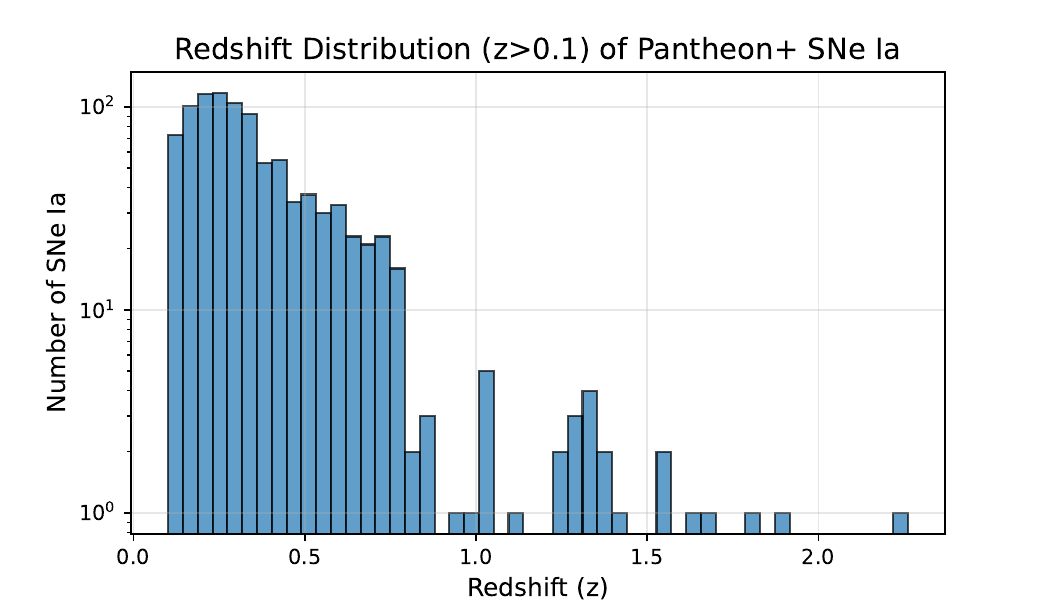}
        \caption{Redshift distribution of the same SNe. The histogram shows the number of supernovae per redshift bin.}
        \label{fig:redshift_density}
    \end{subfigure}
    \caption{(a) Sky distribution and (b) redshift distribution of the Pantheon+ SNe after applying the $z>0.1$ cut. The sample consists of 960 objects.}
    \label{fig:dist}
\end{figure}

\item \textbf{BAO Data:}
We use a compilation of baryon acoustic oscillation (BAO) measurements spanning a wide redshift range, including both isotropic and anisotropic constraints. The low-redshift measurements are given in terms of the volume-averaged distance $D_V/r_d$, while higher-redshift measurements provide constraints on the transverse comoving distance $D_M/r_d$.

The BAO data used in this work are summarized in Table~\ref{tab:bao_data} with their references. These measurements are combined with supernova data to reconstruct the luminosity distance and test the distance duality relation.

\begin{table}[htbp]
\centering
\caption{Compilation of BAO measurements used in this work. The first set corresponds to isotropic measurements of $D_V/r_d$, while the second set provides constraints on the transverse comoving distance $D_M/r_d$.}
\label{tab:bao_data}
\begin{tabular}{ccccc}
\hline
$z_{\rm eff}$ & Observable & Value & Survey & Ref\\
\hline
0.106 & $D_V/r_d$ & $2.976 \pm 0.133$ & 6dFGS& \cite{Beutler2011} \\
0.150 & $D_V/r_d$ & $4.470 \pm 0.168$  & SDSS DR7 & \cite{Ashley2015}\\
0.295 & $D_V/r_d$ & $7.944 \pm 0.075$ &DESI DR2 & \cite{DESI:2025zgx}\\
\hline
0.38 & $D_M/r_d$ & $10.23 \pm 0.17$ & SDSS DR12 BOSS & \cite{Alam2017}\\
0.51 & $D_M/r_d$ & $13.588 \pm 0.167$ & DESI DR2 & \cite{DESI:2025zgx}\\
0.61 & $D_M/r_d$ & $15.61 \pm 0.25$  &SDSS DR12 BOSS &\cite{Alam2017}\\
0.698 & $D_M/r_d$ & $17.65 \pm 0.30$ & eBOSS DR16 & \cite{Bautista2020}\\
0.85 & $D_M/r_d$ & $19.51 \pm 0.41$ & DES Y6& \cite{Abbott2024}\\
0.93 & $D_M/r_d$ & $21.71 \pm 0.28$ & DESI DR1 BAO&\cite{DESI2024}\\
1.317 & $D_M/r_d$ & $27.79 \pm 0.69$ & DESI DR1 BAO &\cite{DESI2024}\\
1.48 & $D_M/r_d$ & $30.69 \pm 0.80$ & eBOSS DR16 Quasar &\cite{Ata2020}\\
2.33 & $D_M/r_d$ & $37.50 \pm 1.10$ &eBOSS DR16 Ly$\alpha$ &\cite{duMasdesBourboux2020}\\
\hline
\end{tabular}
\end{table}

The BAO points are selected from independent surveys and non-overlapping redshift bins, so they are treated as uncorrelated.

\item \textbf{Cosmic Chronometers (CC):}
We use 32 measurements of the Hubble expansion rate $H(z)$ obtained from the cosmic chronometer method \citep{Jimenez_2002}. This technique estimates $H(z) = -[1/(1+z)]\,dz/dt$ using the relative ages of passively evolving galaxies, providing a direct, model-independent measurement of the expansion history. The data are listed in Table~\ref{tab:Hz} with their respective references.
\end{enumerate}
\begin{table}[h!]
\centering
\caption{$H(z)$ measurements with $1\sigma$ uncertainties used in this work.}
\label{tab:Hz}
\setlength{\tabcolsep}{2.5pt}   
\renewcommand{\arraystretch}{0.95} 
\footnotesize  

\begin{tabular}{c c c c c c c c}
\hline\hline
$z$ & $H$ & $\sigma$ & Ref. & $z$ & $H$ & $\sigma$ & Ref.\\
\hline
0.07 & 69.0 & 19.6 & \cite{Zhang:2012mp} & 0.4783 & 83.8 & 10.2 & \cite{Moresco:2016mzx}\\
0.09 & 69 & 12 & \cite{Jimenez:2003iv} & 0.48 & 97.0 & 62.0 & \cite{Stern:2009ep}\\
0.12 & 68.6 & 26.2 & \cite{Zhang:2012mp} & 0.5929 & 107.0 & 15.5 & \cite{Moresco:2012jh}\\
0.17 & 83.0 & 8.0 & \cite{Simon:2004tf} & 0.6797 & 95.0 & 10.5 & \cite{Moresco:2012jh}\\
0.1791 & 78.0 & 6.2 & \cite{Moresco:2012jh} & 0.75 & 98.8 & 33.6 & \cite{Borghi_2022}\\
0.1993 & 78.0 & 6.9 & \cite{Moresco:2012jh} & 0.7812 & 96.5 & 12.5 & \cite{Moresco:2012jh}\\
0.20 & 72.9 & 29.6 & \cite{Zhang:2012mp} & 0.8754 & 124.5 & 17.4 & \cite{Moresco:2012jh}\\
0.27 & 77.0 & 14.0 & \cite{Simon:2004tf} & 0.88 & 90.0 & 40.0 & \cite{Stern:2009ep}\\
0.28 & 88.8 & 36.6 & \cite{Zhang:2012mp} & 0.9 & 117 & 23 & \cite{Simon:2004tf}\\
0.3519 & 85.5 & 15.7 & \cite{Moresco:2012jh} & 1.037 & 133.5 & 17.6 & \cite{Moresco:2012jh}\\
0.3802 & 83 & 13.5 & \cite{Moresco:2016mzx} & 1.3 & 168.0 & 17.0 & \cite{Simon:2004tf}\\
0.4 & 95 & 17 & \cite{Simon:2004tf} & 1.363 & 160.0 & 33.8 & \cite{Moresco:2015cya}\\
0.4004 & 79.9 & 11.4 & \cite{Moresco:2016mzx} & 1.43 & 177.0 & 18.0 & \cite{Simon:2004tf}\\
0.4247 & 90.4 & 12.8 & \cite{Moresco:2016mzx} & 1.53 & 140.0 & 14.0 & \cite{Simon:2004tf}\\
0.4497 & 96.3 & 14.4 & \cite{Moresco:2016mzx} & 1.75 & 202.0 & 40.0 & \cite{Simon:2004tf}\\
0.47 & 89.0 & 49.6 & \cite{Ratsimbazafy:2017vga} & 1.965 & 186.5 & 50.6 & \cite{Moresco:2015cya}\\
\hline
\end{tabular}

\end{table}
This combination of data sets, SNe Ia for luminosity distances, BAO for angular diameter distances, and CC for the expansion history, allows a self-consistent, model-independent test of the distance duality relation and its possible anisotropy.

\subsection{Reconstruction of BAO distances using Gaussian processes}

Baryon acoustic oscillation (BAO) measurements are available only at discrete redshifts, while the supernova sample spans a nearly continuous redshift distribution. To enable a consistent comparison, we reconstruct a smooth luminosity distance function from the BAO data using Gaussian Process Regression (GPR) \citep{Rasmussen2006,Aigrain2023}. GPR provides a non‑parametric, model‑independent interpolation that returns both the mean function and its full covariance. We implement the GPR using the \texttt{GaussianProcessRegressor} module of \texttt{scikit-learn} \citep{Pedregosa2011}.

\subsubsection{BAO data and conversion to luminosity distance}

Our BAO compilation consists of measurements expressed either as the transverse comoving distance $D_M/r_d$ or as the spherically averaged distance $D_V/r_d$. The sound horizon $r_d$ is treated as a free nuisance parameter, we adopt a Gaussian prior from Planck 2018 \citep{planck2018}, $r_d = 147.09 \pm 0.26\ \mathrm{Mpc}$. This prior anchors the BAO scale to the CMB while allowing the data to determine the absolute scale in combination with the SNIa absolute magnitude $M_B$, for which we consider different cases (see Sec.~\ref{sec:likelihood}).

For a measurement in the form $D_M/r_d$, the angular diameter distance is
\begin{equation}
D_A(z) = r_d\, \frac{(D_M/r_d)}{1+z},
\end{equation}
and the corresponding luminosity distance is $D_L = (1+z)^2 D_A$.

For isotropic BAO measurements $D_V/r_d$, the conversion requires the expansion history $H(z)$. Using the definition of the spherically averaged distance,
\begin{equation}
D_V(z) = \left[ (1+z)^2 D_A^2(z)\, \frac{c z}{H(z)} \right]^{1/3},
\end{equation}
we obtain
\begin{equation}
\frac{D_A(z)}{r_d} = \frac{(D_V/r_d)^{3/2}\, r_d^{1/2} \, \sqrt{H(z)}}{\sqrt{c\,z}\,(1+z)} .
\end{equation}
Therefore, an independent, model‑independent reconstruction of $H(z)$ is required to use the $D_V/r_d$ data.

\subsubsection{Kernel choice and hyperparameter optimisation}
\label{sec:kernel_choice}

In GPR, the covariance between two points $z$ and $z'$ is specified by a kernel $k(z,z')$. We consider two widely used forms in cosmology \citep{Seikel2012},

\begin{itemize}
\item \textbf{Squared Exponential (RBF) kernel}
\begin{equation}
k_{\mathrm{RBF}}(z,z') = \sigma_f^2 \exp\!\left(-\frac{(z-z')^2}{2\ell^2}\right),
\end{equation}

which gives infinitely differentiable, very smooth functions.
\item \textbf{Matern kernel} with half‑integer $\nu = 3/2,5/2,7/2$.
\end{itemize}

In all cases we add a white‑noise component $\sigma_n^2\delta_{ij}$ to account for observational uncertainties.

We perform 5‑fold cross‑validation based on the coefficient of determination $R^2$ to select the most appropriate kernel for each dataset. For the CC $H(z)$ data, the Matern $\nu=3/2$ kernel gives the highest cross‑validation score, reflecting the sparse and mildly irregular nature of these measurements. Higher $\nu$ kernels tend to over‑smooth, while the RBF is too rigid. For the BAO‑derived $D_L^{\mathrm{BAO}}(z)$ data, the RBF kernel is preferred, as expected for the smooth distance-redshift relation, Matern kernels with $\nu\ge5/2$ yield nearly identical results but are slightly disfavoured by the cross‑validation metric.

Consequently we adopt
\[
H(z) \;\rightarrow\; \text{Matern } (\nu=3/2), \qquad
D_L^{\mathrm{BAO}}(z) \;\rightarrow\; \text{RBF kernel}.
\]

For each kernel, the hyperparameters $\{\sigma_f, \ell, \sigma_n\}$ are determined by maximising the log‑marginal likelihood (type‑II maximum likelihood) \citep{Rasmussen2006}. This data‑driven choice ensures an optimal balance between flexibility and stability. To improve numerical stability, we scale the input redshifts to $[0,1]$ and standardise the output distances to zero mean and unit variance before training, the predictions are subsequently rescaled back to physical units. The resulting covariance matrix $C_{\mathrm{BAO}}$ is used in the construction of the CDDR parameter $\eta(z)$ (Sec.~\ref{sec:eta_recons}).

\subsubsection{Reconstruction of $H(z)$ from cosmic chronometers}

\begin{figure}[ht]
\centering
\includegraphics[width=0.7\textwidth]{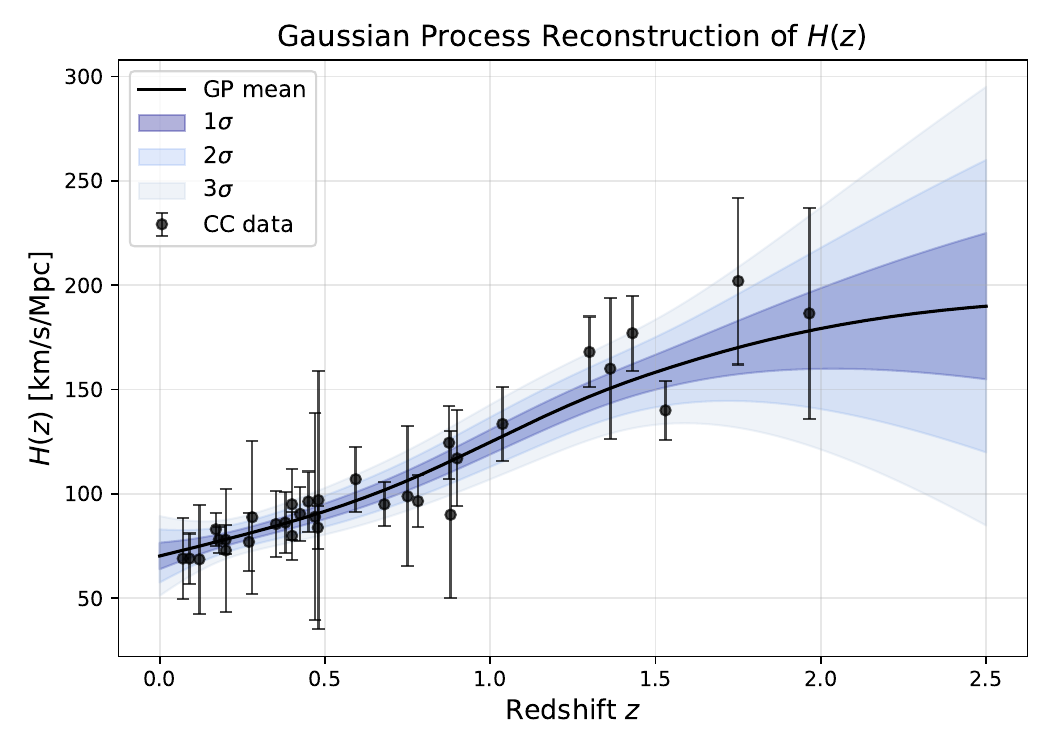}
\caption{Reconstruction of the Hubble parameter $H(z)$ from cosmic chronometer data using GPR. The shaded region indicates the $1\sigma$ uncertainty.}
\label{fig:Hz_reconstruction}
\end{figure}

We reconstruct the Hubble parameter $H(z)$ from 32 cosmic chronometer (CC) measurements using GPR. After testing several kernels via 5‑fold cross‑validation (Sec.~\ref{sec:kernel_choice}), we select the Matern kernel with $\nu=3/2$ for the CC data. The resulting reconstruction, shown in Fig.~\ref{fig:Hz_reconstruction}, is used exclusively for converting $D_V/r_d$ into $D_A(z)$.

\subsubsection{Reconstruction of $D_L^{\mathrm{BAO}}(z)$ from BAO data}

\begin{figure}[ht]
\centering
\includegraphics[width=0.7\textwidth]{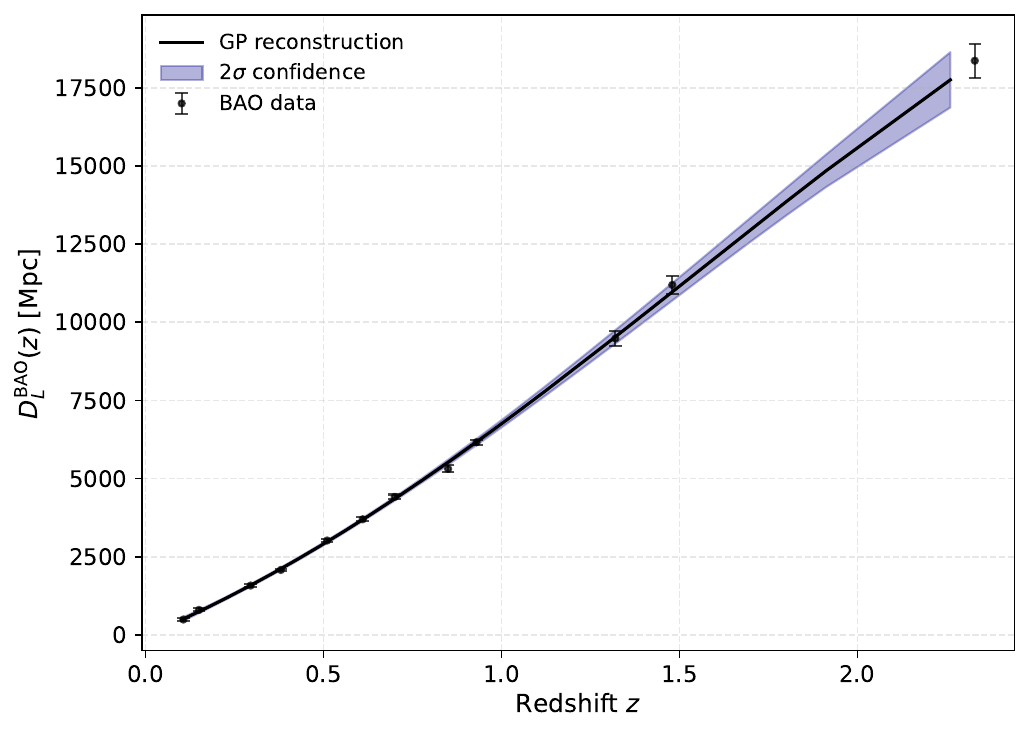}
\caption{Illustrative reconstruction of the BAO‑derived luminosity distance $D_L^{\mathrm{BAO}}(z)$ using GPR, with $r_d=147.09$Mpc and $M_B=-19.385$ fixed. The shaded region shows the $2\sigma$ confidence interval, evaluated at the supernova redshifts.}
\label{fig:DL_reconstruction}
\end{figure}

Using the conversions above, we obtain a set of BAO‑derived luminosity distances $D_L^{\mathrm{BAO}}(z_i)$ at the redshifts of the individual BAO measurements. We then apply a second GPR to reconstruct a smooth function $\bar{D}_L^{\mathrm{BAO}}(z)$ over the entire redshift range covered by the supernova sample ($0.1 < z < 2.3$). For this reconstruction we adopt the RBF kernel. The GPR returns the mean prediction and the full covariance matrix $C_{\mathrm{BAO}}$, the latter is propagated into the subsequent analysis of the CDDR parameter $\eta(z)$. Figure~\ref{fig:DL_reconstruction} shows the reconstructed luminosity distance with its $2\sigma$ confidence band.

\subsection{Reconstruction of $\eta(z)$ and its covariance} \label{sec:eta_recons}

The luminosity distance for Type Ia supernovae is derived from the observed apparent magnitude $m_B$ as
\begin{equation}
D_L^{\rm SN}(z_i) = 10^{\frac{m_{B,i} - M_B - 25}{5}},
\end{equation}
where $M_B$ denotes the absolute magnitude of supernovae, treated as a free nuisance parameter.

The Pantheon+ compilation provides the full covariance matrix $C^{\rm SN}_{ij}$ in terms of the distance modulus. To express the covariance in luminosity distance space, we propagate uncertainties using this transformation,
\begin{equation}
(C^{DL}_{\rm SN})_{ij} =
\left(\frac{\ln 10}{5} D_{L,i}^{\rm SN}\right)
\left(\frac{\ln 10}{5} D_{L,j}^{\rm SN}\right)
C^{\rm SN}_{ij}.
\end{equation}
This consistently accounts for both statistical and systematic uncertainties in the supernova data.

The BAO luminosity distance $D_L^{\rm BAO}(z)$ and its covariance matrix $C^{\rm BAO}_{ij}$ are obtained from the Gaussian Process reconstruction described in the previous subsection. The CDDR observable is then constructed at the redshift of each supernova as
\begin{equation}
\eta_i = \frac{D_{L,i}^{\rm SN}}{D_{L,i}^{\rm BAO}}.
\end{equation}
Crucially, we use the original supernova measurements without smoothing, the resulting $\eta_i$ therefore exhibit scatter that reflects the individual SN uncertainties, and this scatter is fully accounted for by the covariance matrix derived below.

The covariance matrix of $\eta$ is computed through linear error propagation, incorporating contributions from both supernova and BAO uncertainties.

The contribution arising from supernova uncertainties is given by
\begin{equation}
(C_\eta^{\rm SN})_{ij} =
\frac{(C^{DL}_{\rm SN})_{ij}}
{D_{L,i}^{\rm BAO} \, D_{L,j}^{\rm BAO}},
\end{equation}
while the contribution from BAO uncertainties reads
\begin{equation}
(C_\eta^{\rm BAO})_{ij} =
\frac{D_{L,i}^{\rm SN} \, D_{L,j}^{\rm SN}}
{\left(D_{L,i}^{\rm BAO}\right)^2 \left(D_{L,j}^{\rm BAO}\right)^2}
\; C^{\rm BAO}_{ij}.
\end{equation}

Assuming that supernova and BAO measurements are statistically independent, the total covariance matrix is obtained as
\begin{equation}
C_\eta = C_\eta^{\rm SN} + C_\eta^{\rm BAO}.
\end{equation}

This procedure preserves the full covariance structure of both datasets and ensures a consistent propagation of uncertainties in the reconstruction of $\eta(z)$, while avoiding any artificial smoothing of the supernova data that could bias the dipole analysis.

\begin{figure}[t]
\centering
\includegraphics[width=0.7\textwidth]{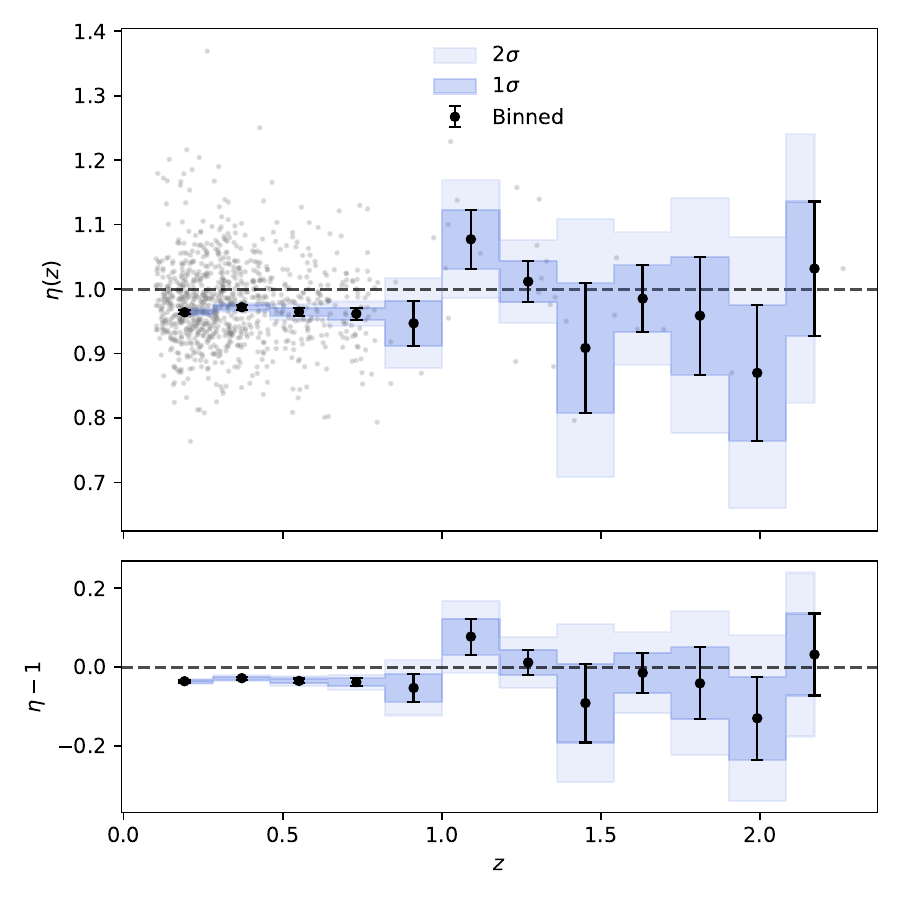}
\caption{Reconstructed $\eta(z)$ as a function of redshift (top panel) and the corresponding residuals $\eta-1$ (bottom panel), obtained from the combined supernova and BAO data. Gray points show individual supernova measurements, while black points represent inverse-variance weighted binned values used for visualization. The shaded regions indicate the $1\sigma$ and $2\sigma$ uncertainties derived from the reconstructed covariance. The dashed lines denote the standard CDDR prediction, $\eta=1$. This figure illustrates the data-driven reconstruction of $\eta(z)$ prior to the likelihood analysis.}
\label{fig:eta_reconstruction}
\end{figure}

In Fig.~\ref{fig:eta_reconstruction} we show the reconstructed $\eta(z)$ and its associated covariance for fixed value $M_B=-19.385$ and $r_d=147.09$. The reconstruction is performed in a model-independent manner, and the binning is introduced only for visualization. The statistical analysis is performed using the full dataset and its associated covariance matrix. Binning the data would lead to a loss of information and a non-trivial modification of the covariance structure and is therefore not suitable for parameter inference.

\subsection{Likelihood analysis} \label{sec:likelihood}

We constrain the model parameters by minimising the total chi‑square function
\begin{equation}
\chi^2 = \bigl(\boldsymbol{\eta} - \boldsymbol{\eta}_{\rm model}\bigr)^{\mathsf{T}} \mathbf{C}^{-1} \bigl(\boldsymbol{\eta} - \boldsymbol{\eta}_{\rm model}\bigr) + \chi^2_{\rm prior},
\end{equation}
where $\boldsymbol{\eta}$ is the vector of reconstructed CDDR values at the supernova redshifts, $\boldsymbol{\eta}_{\rm model}$ is the corresponding model prediction, and $\mathbf{C}$ is the total covariance matrix constructed in Sect.~\ref{sec:eta_recons}. The term $\chi^2_{\rm prior}$ incorporates external Gaussian priors.

\paragraph{Prior on the sound horizon $r_d$.}
We adopt a Gaussian prior from Planck 2018 \citep{planck2018}:
\begin{equation}
\chi^2_{r_d} = \left(\frac{r_d - r_d^{\rm fid}}{\sigma_{r_d}}\right)^2,
\end{equation}
with $r_d^{\rm fid} = 147.09\ \mathrm{Mpc}$ and $\sigma_{r_d} = 0.26\ \mathrm{Mpc}$.

\paragraph{Priors on the supernova absolute magnitude $M_B$.}
We consider three distinct cases to examine the impact of the SN calibration on the CDDR test:

\begin{itemize}
\item \textbf{Case A (free $M_B$):} $M_B$ is treated as a free parameter with a uniform (flat) prior over the physically plausible range $[-20,-18]$.
\item \textbf{Case B (DB23 prior):} A model‑independent Gaussian prior from \cite{Dinda_2023} is applied,
\begin{equation}
\chi^2_{M_B} = \left(\frac{M_B - M_B^{\rm DB23}}{\sigma_{M_B}}\right)^2,
\end{equation}
with $M_B^{\rm DB23} = -19.385$ and $\sigma_{M_B} = 0.052$.
\item \textbf{Case C (SH0ES prior):} A Gaussian prior from the SH0ES distance ladder \citep{Riess_2022} is applied,
\begin{equation}
\chi^2_{M_B} = \left(\frac{M_B - M_B^{\rm SH0ES}}{\sigma_{M_B}}\right)^2,
\end{equation}
with $M_B^{\rm SH0ES} = -19.253$ and $\sigma_{M_B} = 0.027$.
\end{itemize}

The likelihood is then $\mathcal{L} \propto \exp(-\chi^2/2)$. We sample the posterior distribution using the affine‑invariant ensemble sampler \texttt{emcee} \citep{emcee} to obtain constraints on the model parameters $\Theta = \{\eta_1, D_x, D_y, D_z, r_d, M_B\}$. From the dipole vector $\mathbf{D} = (D_x, D_y, D_z)$ we compute the amplitude $A = |\mathbf{D}|$ and the direction in Galactic coordinates $(l, b)$. Convergence of the Markov chains is assessed with the Gelman-Rubin statistic, requiring $R < 1.05$ for all parameters.

\subsection{Null test and statistical significance of the dipole} \label{sec:null_test}

To assess the statistical significance of the inferred dipole, we construct null realizations under the hypothesis of an isotropic universe. The goal is to determine whether the observed dipole amplitude can arise from statistical fluctuations and survey geometry alone.

We adopt the dipole parametrization
\begin{equation}
\eta(z,\hat{n}) = (1+\eta_1 z)\left(1 + \hat{n}\cdot \mathbf{D}\right),
\end{equation}
where $\mathbf{D} = (D_x, D_y, D_z)$ is the dipole vector, and the amplitude is defined as $A = |\mathbf{D}|$.

For each samples, we estimate the dipole parameters by minimizing the $\chi^2$ function,
\begin{equation}
\chi^2 = (\boldsymbol{\eta} - \boldsymbol{\eta}_{\rm model})^{\rm T} \mathbf{C}^{-1} (\boldsymbol{\eta} - \boldsymbol{\eta}_{\rm model}),
\end{equation}
where $\mathbf{C}$ is the covariance matrix. In the null tests, we determine the best-fit dipole amplitude $A$ using a numerical optimizer (scipy \texttt{minimize}), which provides an efficient estimate of the maximum-likelihood solution.

We construct two types of null realizations,
\begin{itemize}
\item \textbf{Survey-preserving null:} The supernova observables ($m_b$) are randomly shuffled within redshift bins, while keeping the sky positions fixed. This preserves both the redshift distribution and the survey selection function, allowing us to quantify the dipole induced by non-uniform sky coverage.

\item \textbf{Isotropic null:} The sky positions are replaced by isotropically distributed random unit vectors, while the observables are shuffled. This removes the effect of survey geometry and provides an estimate of the intrinsic statistical noise of the estimator.
\end{itemize}

The statistical significance of the observed dipole is quantified by the $p$-value, defined as the fraction of null realizations with dipole amplitude larger than the observed value,
\begin{equation}
p = \frac{1}{N} \sum_{i=1}^{N} \mathbb(A_i \geq A_{\rm obs}).
\end{equation}

This procedure enables a direct comparison between the observed dipole and the distribution expected from isotropic models, while explicitly accounting for survey geometry and statistical fluctuations.

\section{Results and Discussion}

\begin{figure}[ht]
\centering
\includegraphics[width=0.7\textwidth]{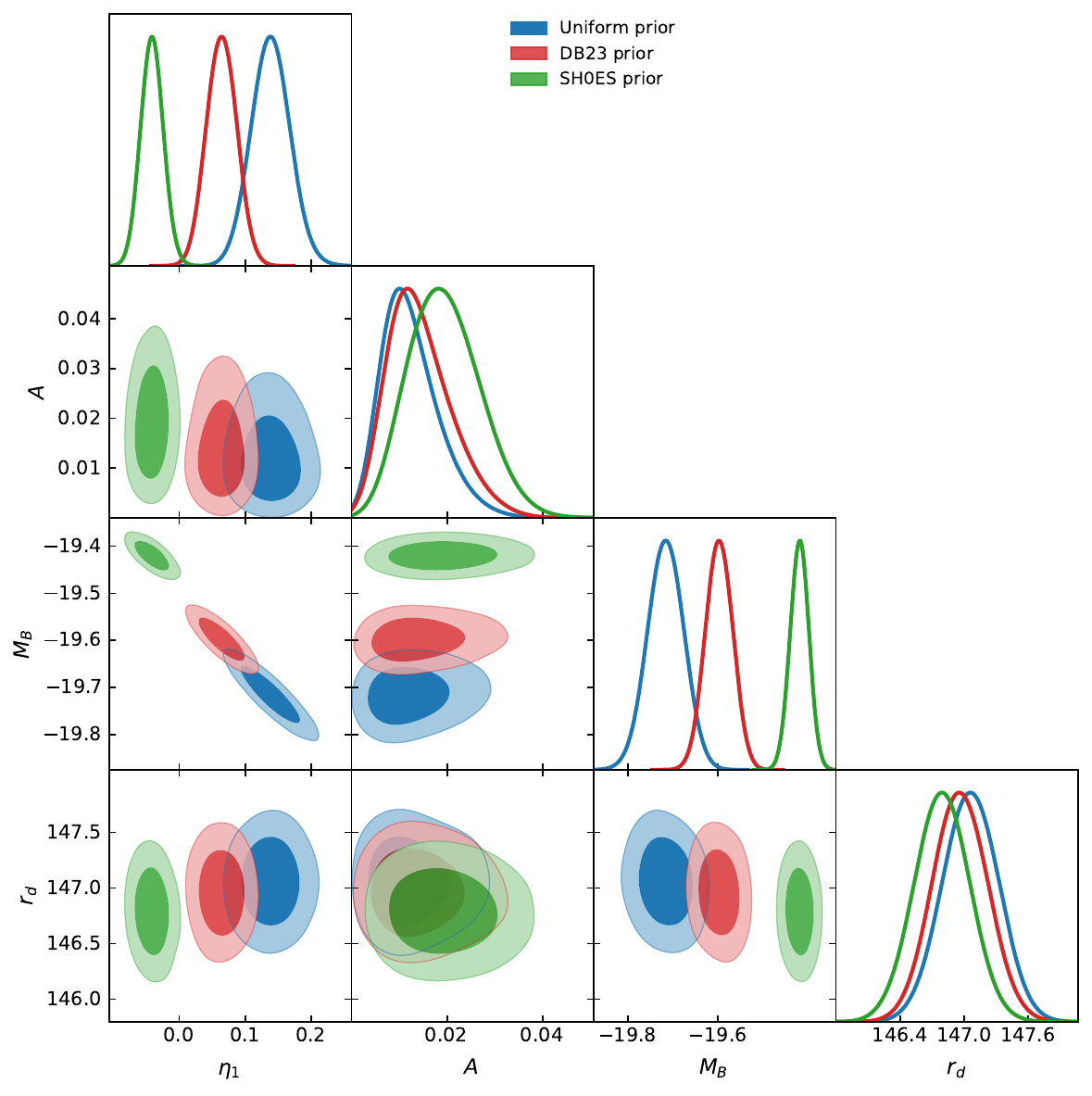}
\caption{Marginalized posterior distributions for the anisotropic DDR parameters for three choices of the supernova absolute magnitude $M_B$ prior. Contours correspond to 68\% and 95\% confidence regions. The dipole amplitude $A$ is consistent across all cases, while $\eta_1$ exhibits a clear dependence on the assumed calibration due to its degeneracy with $M_B$.}
\label{fig:corner_all}
\end{figure}

We analyse the three calibration cases described in Sec.~\ref{sec:likelihood} (free $M_B$, DB23 prior and SH0ES prior.
All analyses use the same BAO data, SN sample, and dipole model $\eta(z,\hat{n}) = (1+\eta_1 z)(1+\hat{n}\cdot\mathbf{D})$. The resulting marginalized constraints are summarised in Table~\ref{tab:results_compare}, and the corresponding posterior distributions are shown in Fig.~\ref{fig:corner_all}.

\begin{table}[ht]
\centering
\caption{Marginalized 68\% credible intervals for the model parameters obtained from the MCMC analysis under three different priors on the supernova absolute magnitude $M_B$: uniform prior (Case~A), DB23 Gaussian prior (Case~B), and SH0ES prior (Case~C). In each case, the sound horizon $r_d$ is constrained by a Planck prior. The table lists the isotropic evolution $\eta_1$, the posterior mean dipole amplitude $A$, the 95\% upper bound on the amplitude $A_{95}$, the nuisance parameters $M_B$ and $r_d$, and the dipole direction $(\ell, b)$.}
\label{tab:results_compare}
\begin{tabular}{lccc}
\hline
Parameter & Case A (uniform $M_B$) & Case B (DB23 prior) & Case C (SH0ES prior)\\
\hline
$\ell$ & $238^{+100}_{-200}$ & $263^{+100}_{-20}$ & $302^{+60}_{-10}$\\
$b$ & $20^{+20}_{-20}$ & $21^{+20}_{-20}$ & $21^{+10}_{-20}$ \\
$\eta_1$ & $0.138 \pm 0.029$ & $0.064 \pm 0.022$ & $-0.041 \pm 0.017$\\
$A$ & $0.0122^{+0.0041}_{-0.0070}$ & $0.0141^{+0.0048}_{-0.0078}$ & $0.0193^{+0.0067}_{-0.0082}$ \\
$A_{95}$ & $0.025$ & $0.029$ & $0.034$\\
$M_B$ & $-19.716 \pm 0.040$ & $-19.598 \pm 0.030$ & $-19.420 \pm 0.021$\\
$r_d$ & $147.06 \pm 0.26$ & $146.96 \pm 0.26$ & $146.79 \pm 0.26$\\
\hline
\end{tabular}
\end{table}

The dipole amplitude is found to be small and consistent across all calibration choices, with $A \sim 0.01$ in all cases. This demonstrates that the constraint on anisotropy is stable and largely independent of the assumed supernova calibration. The dipole direction remains poorly constrained, with broad posteriors in longitude and a latitude consistent with zero, indicating no statistically significant preferred axis.

In contrast, the isotropic evolution parameter $\eta_1$ exhibits a clear dependence on the assumed prior on $M_B$. In Case~A, where the absolute scale is unconstrained, we obtain $\eta_1 = 0.138 \pm 0.029$, corresponding to a $\sim4.8\sigma$ deviation from the standard distance duality relation. When external calibration is imposed (Cases~B and~C), the value of $\eta_1$ shifts significantly, reflecting its strong degeneracy with $M_B$. The DB23 prior yields an intermediate value $\eta_1 = 0.064 \pm 0.022$, while the SH0ES prior drives $\eta_1$ to slightly negative values. This behavior is expected, as changes in the absolute magnitude rescale the luminosity distance and are partially absorbed by the redshift evolution term. These results demonstrate that while the isotropic evolution is sensitive to calibration, the anisotropy constraint is not.

We also repeat the MCMC analysis using a more restrictive redshift cut, $z > 0.15$, thereby excluding the low-redshift supernovae that may be more susceptible to local inhomogeneities and peculiar velocity effects. We find that the constraints on both the dipole amplitude $A$ and the isotropic evolution parameter $\eta_1$ remain consistent within uncertainties with the baseline analysis using $z > 0.1$. This demonstrates that our results are not driven by low-redshift systematics and are stable against reasonable variations in the sample selection.

As a consistency check, we repeat the analysis assuming the isotropy by setting $\mathbf{D}=0$. We find that the constraints on $\eta_1$, $M_B$, and $r_d$ remain consistent with those obtained in the anisotropic analysis across all prior choices. This is expected because of the small amplitude of the dipole. Hence, the anisotropic component does not significantly affecting the estimation of isotropic parameters.

To assess the statistical significance of the dipole, we perform null tests using 500 mock samples, as discussed in Sec.~\ref{sec:null_test}. For each sample, we determine the dipole amplitude by minimizing the $\chi^2$ function using a numerical optimizer (scipy \texttt{minimize}), rather than performing a full MCMC analysis. We consider the $\chi^2$ function,
\begin{equation}
    \chi^2=\chi^2_{\eta}+ \chi^{2}_{r_d}.
\end{equation}
This approach is sufficient for estimating the best-fit dipole amplitude in each mock and enables an efficient construction of the null distribution, while the full posterior inference is reserved for the analysis of the real data. For the real data, $\chi^2$ minimization give $A_{real} \simeq 0.008$.

In the first test, we generate the mock samples by randomly shuffling the supernova observables ($m_B$) within redshift bins while keeping their sky positions fixed. The resulting distribution of dipole amplitudes peaks at $A \sim 0.048$, indicating a false positive signal induced by the non-uniform sky coverage. In the second test, we replace the observed sky positions with isotropically distributed random vectors and shuffle the supernova magnitude ($m_B$) within redshift bins to generate the mock samples. The dipole amplitude from these mock samples peaks around $A \sim 0.017$. Which shows that in the absence of a true anisotropic signal, the best-fit dipole picks the statistical fluctuations in the data, leading to a non-zero amplitude. The actual dipole we observe from real data, $A_{real}$, is much smaller than these mock distributions. The p-value comes out close to one, meaning the data show no real sign of anisotropy in the CDDR.

We quantify the dipole significance using posterior obtained from real data and from the null distributions. From the marginalised posterior, we obtain a 95\% upper bound $A < 0.025$. For comparison, the 95th percentile of the isotropic null distribution is $A_{95}^{\rm iso} = 0.03$, while the corresponding value obtained using the real survey geometry is $A_{95}^{\rm real} = 0.06$.

Our observed limit is smaller than both thresholds, confirming that the dipole is fully consistent with isotropy. The large gap between $A_{95}^{\rm iso}$ and $A_{95}^{\rm real}$ also highlights how uneven sky coverage can artificially inflate the dipole signal. Thus, the apparent anisotropy arises mainly from statistical noise and survey geometry, not from any genuine directional dependence in the universe.

\begin{figure}[ht]
\centering
\includegraphics[width=0.7\textwidth]{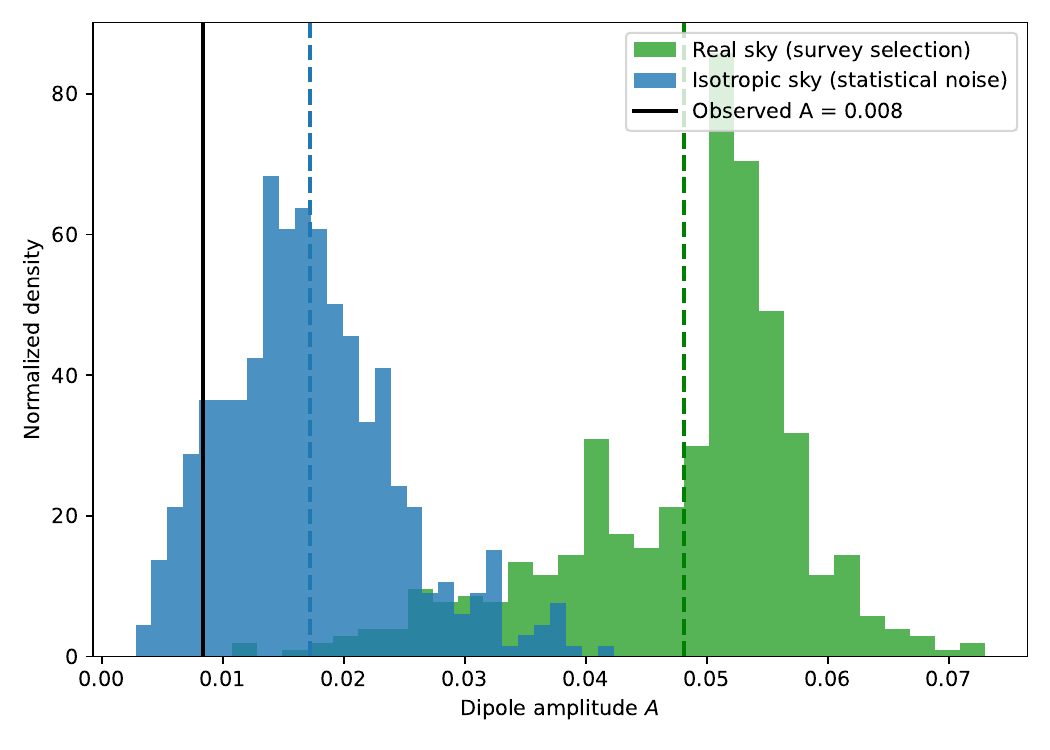}
\caption{Distribution of dipole amplitudes obtained from null samples. The green histogram corresponds to shuffled supernova $m_b$ with the real sky positions, capturing the effect of the survey selection function. The blue histogram shows the same procedure applied to isotropically distributed sky positions, representing the statistical noise. The vertical black line indicates the observed dipole amplitude.}
\label{fig:null}
\end{figure}

In summary, our joint analysis of Pantheon+ supernovae and BAO data finds no evidence for a dipole anisotropy in the distance duality relation. The dipole amplitude is constrained to be at the percent level and is fully consistent with statistical fluctuations. While the isotropic evolution parameter $\eta_1$ does depend on how the supernova distances are calibrated, the absence of anisotropy is a calibration‑independent result. Future wide‑area surveys with better sky coverage will be essential to tighten these limits further and to probe anisotropic effects beyond the current noise and survey selection limits.

\section{Conclusion}

In this work, we have performed a model‑independent test of anisotropy in the cosmic distance duality relation (CDDR). We combined the Pantheon+ supernova sample (redshift $z>0.1$) with baryon acoustic oscillation (BAO) data. The angular diameter distance was reconstructed using Gaussian processes, allowing us to obtain $\eta(z)$ without assuming any specific cosmological model. We then extended the standard CDDR to include a dipole modulation and constrained the parameters using a full covariance‑based likelihood, considering different choices for the supernova absolute magnitude $M_B$.

We find no evidence for a statistically significant dipole. The inferred dipole amplitude is small and fully consistent with isotropy. From the marginalized posterior distribution, we obtain a 95\% upper limit of $A < 0.025$ for our baseline case (uniform prior on $M_B$), with similar bounds obtained under the other calibration assumptions. This shows that the anisotropy constraint is robust against the choice of distance scale calibration.

To assess the statistical significance, we constructed null samples that preserve both the redshift distribution and the actual survey selection function. The observed dipole amplitude lies below the level expected from isotropic realisations and is well below the threshold induced by the non‑uniform sky coverage, i.e. $A_{\text{obs}} < A_{\text{iso}} < A_{\text{survey}}$. The large difference between the isotropic and the survey‑preserving null distributions highlights how survey geometry can artificially inflate the apparent dipole signal. Overall, our results indicate that the observed anisotropy is consistent with statistical fluctuations and selection effects, with no evidence for an intrinsic directional dependence.

We have also verified that including the dipole component does not noticeably change the constraints on the isotropic parameters $\eta_1$, $M_B$, and $r_d$. This confirms that the anisotropic contribution is subdominant and effectively decoupled from the isotropic sector.

In summary, our analysis provides no evidence for a violation of statistical isotropy in the CDDR. More broadly, this work demonstrates the importance of carefully accounting for survey selection effects and covariance structure when searching for anisotropic signals in cosmological data. The methodology developed here can be directly applied to future surveys with improved sky coverage and statistical precision, enabling even more stringent tests of the cosmological principle.

\section{Acknowledgment}

G.K. acknowledges financial support from Vellore Institute of Technology through its Seed Grant (No. SG20230035, 2023).

\bibliographystyle{JHEP}
\bibliography{DDR}

\end{document}